# Structural Analysis of Population Graphs

Kimberly Ayers, Maxwell Kooiker

**Abstract**

The format of graphing algorithms for genomic data has been a debate in recent biotechnology. In this paper, we discuss the construction of "population graphs" using said genomic data. We first examine the GENPOFAD distance measurement, developed by Joly et. al., and prove that this constitutes a metric function. We develop an algorithm to construct graphs to visualize the relationships between individuals in a population. We then provide a statistical analysis of these simulated population graphs, and show that they are distinct from randomly generated graphs, and also show differences from small-world graphs.

## Introduction

Networks are a convenient way to visualize data because they clearly show relationships and interactions between entities, making complex systems easier to understand at a glance. By representing items as nodes and their connections as edges, networks can reveal patterns such as clusters, hubs, or isolated points that might be hidden in a spreadsheet or table. This visual structure allows for quick identification of key players, bottlenecks, and highly connected components, which is especially valuable in fields like social science, biology, and computer science. Additionally, the flexibility of network diagrams means they can be scaled to represent both small, simple datasets and large, intricate systems, making them a versatile tool for analysis and communication.

Visualizing genetic data through graphs provides a powerful means to explore, analyze, and communicate the complex relationships inherent in biological systems. Graph-based approaches represent entities such as genes, proteins, or genetic variants as nodes, and their functional, regulatory, or evolutionary relationships as edges. This structure enables researchers to capture multidimensional data in an intuitive and interpretable form, revealing patterns such as gene co-expression modules, regulatory hierarchies, and network motifs that may be difficult to detect using tabular or linear representations (Palla et al., 2005). Advances in computational biology have further enhanced the utility of graph visualization, with interactive and scalable tools allowing researchers to integrate heterogeneous data sources, from genomic sequences to phenotypic traits, into unified network models (Pavlopoulos, et al., 2011). By leveraging the strengths of graph theory and modern visualization techniques, genetic network diagrams facilitate hypothesis generation, support comparative analysis across datasets, and promote a systems-level understanding of genetic function and organization.

In close-knit communities, graphs representing relationships are often expected to exhibit **small-world** properties, characterized by high clustering and short average path lengths. This arises because individuals in such communities tend to form tightly interconnected groups, where friends of friends are likely to also be friends, producing densely connected clusters. At the same time, a few connections that span different subgroups—such as acquaintances through work, school, or family—create shortcuts that dramatically reduce the number of steps needed to connect any two individuals in the network. This combination of local cohesion and global reach mirrors the small-world phenomenon described in (Watts & Strogatz, 1998) and is observed in many real-world social, biological, and information networks. Such structure facilitates rapid information flow, reinforces social cohesion, and can make communities more resilient to the loss of individual connections.

In this paper, we aim to construct graphs of communities of individuals and map their genetic relationships. We generate communities of individuals using algorithms presented in (Adrion, et al., 2020) and (Gutenkunst, Hernandez, Williamson, & Bustamante, 2009), and use genetic distance measures introduced in (Joly, Bryant, & Lockhart, 2015). We prove that the genetic distance measure known as genpofad satisfies the definitions of a metric, allowing for a more sophisticated topological analysis of the genetic space. Finally, we show that the graphs generated in this space are distinct from randomly generated graphs, and also at times seem to differ from small-world networks.

## Background and Definitions

We will use this section to introduce the genetic and graph theoretic terminology we will use throughout this article.

**Definition 1**. *A nucleotide site is an entry of DNA consisting of either Adenine (A), Guanine (G), Thymine (T), or Cytosine (C).*

**Definition 2**. *A nucleotide sequence is a string of nucleotide sites.*

**Definition 3**. *A locus (plural loci) is a specific location in genetic code on the region of a chromosome. Since chromosomes come in pairs in humans, each loci is considered as a pair of nucleotide sequences.*

**Definition 4**. *An allele is the form a gene takes at a particular locus or nucleotide site.*

# Graphical Structures in a Genomic Context

In order to construct a graph representing the genetic relationships between a population, we first must have some measure of distance between two nucleotide sequences, located at the same locus in two different individuals. Once we have a distance measure, we can construct a graph by representing individuals as nodes and creating an edge between individuals $i$ and $j$ (where $i \neq j$ if the distance between $i$ and $j$'s nucleotide sequences falls under a certain cutoff value. (For a further discussion of choice of cutoff value, please see section 3.3.1.

## Genetic Distance between Individuals

In , Joly et. al. introduce four new methods for measuring and calculating the genetic distance between individuals from sequence data: MATCHSTATE, GENPOFAD, MRCA, and NEI. All four methods are calculated at the level of nucleotides, thus it requires information of the nucleotide present at a given position for an individual. Additionally, with all four methods, all distances will fall between 0 and 1, and the distance between an individual and themselves will always be 0. The distance measures also take polyploidy into account, in particular the diploid nature of the human genome.

Joly et. al. found that, in most cases, in terms of genetic distance accuracy, GENPOFAD performed the best of their proposed distances, and therefore, will be taken to be the standard genetic distance between individuals at a particular locus in this work. GENPOFAD works by considering a particular nucleotide state indexed by $i$, which appears in two copies in each human. Thus for an individual, a nucleotide state is a string of 2 characters, built from the alphabet $\{A, C, G, T\}$. These copies may be the same (homozygote) (eg, the string "$AA$") or differ (heterozygote) (eg, the string ("$AG$"). Using Joly et. al.'s notation, let $S_X^i$ be the set of nucleotides observed at site $i$ for individual $X$, and let $|S_X^i|$ be the size of that set. Note that $|S_X^i|$ will always be either 1 or 2. For example, suppose individual $X$ is homozygous AA at site $i$; then $S_X^i = \{A\}$ and $|S_X^i| = 1$. The GENPOFAD distance at site $i$ between individuals $X$ and $Y$ is then calculated as

$$\text{GENPOFAD}_{XY}^i := 1 - \frac{|S_X^i \cap S_Y^i|}{\max\{|S_X^i|, |S_Y^i|\}}.$$

This can then be extended to a measure of distance across multiple nucleotide sites within a single locus by averaging:

$$\text{GENPOFAD}_{XY} = \frac{1}{n} \sum_{i=1}^{n} \text{GENPOFAD}_{XY}^i$$

where $n$ is the number of nucleotide sites.

We claim that these GENPOFAD distances form a metric on the set of pairs of strings of a fixed length made up of the alphabet $\{A, C, G, T\}$. Formally, a metric on a set $M$ is a function $d: M \times M \to \mathbb{R}$ satisfying:

1. the distance from an element to itself is always zero: $d(x, x) = 0$ for all $x \in M$

2. the distance between two distinct points is always positive: $d(x, y) > 0$ if and only if $x \neq y$

3. distance is symmetric; that is, the distance from a point $x$ to a point $y$ is the same as the distance from $y$ to $x$: $d(x, y) = d(y, x)$ for all $x, y \in M$

4. the triangle inequality holds: $d(x, z) \leq d(x, y) + d(y, z)$ for all $x, y, z \in M$.

We first claim that the GENPOFAD distance forms a metric on individual nucleotide sites.

**Lemma 5.** *GENPOFAD distance has symmetry on individual nucleotide sites:*

$$\text{genpofad}^i_{XY} = \text{genpofad}^i_{YX}$$

*Proof.*

$$\begin{aligned}
\text{GENPOFAD}^i_{XY} &= 1 - \frac{|S^i_X \cap S^i_Y|}{\max\{|S^i_X|, |S^i_Y|\}} \\
&= 1 - \frac{|S^i_Y \cap S^i_X|}{\max\{|S^i_Y|, |S^i_X|\}} \\
&= \text{GENPOFAD}^i_{YX}.
\end{aligned}$$

By symmetry of the intersection of sets, and the indifference of reordering a set. ∎

**Lemma 6.** *The (GENPOFAD) distance between two nodes is $0$ if and only if the two nodes are equal.*

*Proof.*

(=>) Let there be two nodes whose distance between one another is 0. Then we have:

$$\text{GENPOFAD}^i_{XY} = 0$$
$$\implies 1 - \frac{|S^i_X \cap S^i_Y|}{\max\{|S^i_X|, |S^i_Y|\}} = 0$$
$$\frac{|S^i_X \cap S^i_Y|}{\max\{|S^i_X|, |S^i_Y|\}} = 1$$
$$|S^i_X \cap S^i_Y| = \max\{|S^i_X|, |S^i_Y|\}.$$

Without loss of generality (by symmetry) let $|S^i_X| \geq |S^i_Y|$. Then:

$$|S^i_X \cap S^i_Y| = |S^i_X|.$$

Since $(S^i_X \cap S^i_Y) \subseteq S^i_X$, and since all sets are finite, we determine that $(S^i_X \cap S^i_Y) = S^i_X$.

However, notice that $(S^i_X \cap S^i_Y) = S^i_X \subseteq S^i_Y$ as well, so $|S^i_X| \leq |S^i_Y|$. Recall that we said $|S^i_X| \geq |S^i_Y|$. Thus, $|S^i_X| = |S^i_Y|$. We conclude that $S^i_X = S^i_Y$, since $S^i_X \subseteq S^i_Y$.

(<=) Notice that:

$$\text{GENPOFAD}^i_{XX} = 1 - \frac{|S^i_X \cap S^i_X|}{\max\{|S^i_X|\}}$$
$$= 1 - \frac{|S^i_X|}{|S^i_X|}$$
$$= 1 - 1$$
$$= 0.$$

**Lemma 7.** *GENPOFAD satisfies the triangle inequality:*

$$\text{genpofad}^i_{XZ} + \text{genpofad}^i_{ZY} \leq \text{genpofad}^i_{XY}$$

*That is,*

$$\left(1 - \frac{|S_X^i \cap S_Z^i|}{\max\{|S_X^i|, |S_Z^i|\}}\right) + \left(1 - \frac{|S_Z^i \cap S_Y^i|}{\max\{|S_Z^i|, |S_Y^i|\}}\right) \geq 1 - \frac{|S_X^i \cap S_Y^i|}{\max\{|S_X^i|, |S_Y^i|\}}.$$

*Proof.* The proof will be done in cases regarding the size of $S_X^i \cap S_Y^i$. Notice that these will be integers between (and including) 0 and 2. We will consider three cases, $S_X^i \cap S_Y^i = 0$, $S_X^i \cap S_Y^i = 1$, and $S_X^i \cap S_Y^i = 2$.

Case 1: $|S_X^i \cap S_Y^i| = 0$. If $S_X^i$ and $S_Y^i$ are disjoint, then Equation [eqn1] becomes

$$\left(1 - \frac{|S_X^i \cap S_Z^i|}{\max\{|S_X^i|, |S_Z^i|\}}\right) + \left(1 - \frac{|S_Z^i \cap S_Y^i|}{\max\{|S_Z^i|, |S_Y^i|\}}\right) \geq 1.$$

If $|S_X^i \cap S_Z^i| = 0$:

$$\begin{aligned} 1 - \frac{|S_X^i \cap S_Z^i|}{\max\{|S_X^i|, |S_Z^i|\}} + 1 - \frac{|S_Z^i \cap S_Y^i|}{\max\{|S_Z^i|, |S_Y^i|\}} &= 1 - 0 + 1 - \frac{|S_Z^i \cap S_Y^i|}{\max\{|S_Z^i|, |S_Y^i|\}} \\ &\geq 1 + (1 - 1) \\ &= 1 - 0 \\ &= 1 - \frac{|S_X^i \cap S_Y^i|}{\max\{|S_X^i|, |S_Y^i|\}}. \end{aligned}$$

A similar argument holds if $|S_Y^i \cap S_Z^i| = 0$. If both $|S_X^i \cap S_Z^i|$ and $|S_X^i \cap S_Z^i|$ are nonzero, then it must be that $|S_X^i \cap S_Z^i| = |S_X^i \cap S_Z^i| = 1$, and $|S_X^i| = |S_Y^i| = |S_Z^i| = 2$ since $|S_X^i \cap S_Z^i| = 0$. In this case,

$$1 - \frac{|S_X^i \cap S_Z^i|}{\max\{|S_X^i|, |S_Z^i|\}} + 1 - \frac{|S_Z^i \cap S_Y^i|}{\max\{|S_Z^i|, |S_Y^i|\}} = 1 - \frac{1}{2} + 1 - \frac{1}{2} = 1.$$

Case 2: $|S_X^i \cap S_Y^i| = 1$. In this case, if $|S_Y^i \cap S_Z^i| = 0$, then Equation [eqn1] becomes

$$2 - \frac{|S_X^i \cap S_Z^i|}{\max\{|S_X^i|, |S_Z^i|\}} \geq 1 - \frac{1}{\max\{|S_X^i|, |S_Y^i|\}},$$

which is equivalent to

$$\frac{|S_X^i \cap S_Z^i|}{\max\{|S_X^i|, |S_Z^i|\}} \leq 1 + \frac{1}{\max\{|S_X^i|, |S_Y^i|\}},$$

which holds since $\frac{|S_X^i \cap S_Z^i|}{\max\{|S_X^i|, |S_Z^i|\}} \leq 1$. If $|S_Z^i \cap S_Y^i| = 1$, then $|S_Z^i| = 2$, as $S_Z^i$ must have exactly one distinct element in each set $S_X^i$ and $S_Y^i$. Then equation [eqn1] becomes

$$1 - \frac{1}{2} + 1 - \frac{1}{2} \geq 1 - \frac{|S_X^i \cap S_Y^i|}{\max\{|S_X^i|, |S_Y^i|\}}$$

which holds. Lastly, we can not have $|S_Z^i \cap S_Y^i| = 2$ as then $S_Z^i = S_Y^i$, but $|S_X^i \cap S_Z^i| = 1 \neq 0 = |S_X^i \cap S_Y^i|$. By this contradiction, this case cannot exist.

Case 3: $|S_X^i \cap S_Y^i| = 2$. This implies $S_X^i = S_Y^i$, and that $|S_X^i| = |S_Y^i| = 2$. Equation [eqn1] becomes

$$\left(1 - \frac{|S_X^i \cap S_Z^i|}{2}\right) + \left(1 - \frac{|S_Z^i \cap S_Y^i|}{2}\right) \geq 1 - \frac{2}{2}$$

which simplifies to

$$2 - |S_X^i \cap S_Z^i| \geq 0$$

which holds since $|S_X^i \cap S_Z^i| \leq 2$. Thus, in all 3 cases, the triangle inequality is satisfied. ∎

**Theorem 8**. *As defined above, GENPOFAD distances form a metric.*

*Proof.* This immediately follows from combining Lemmas 14, 15, and 16 and noting that GENPOFAD always gives distances between 0 and 1 (and thus is always non-nonegative). ∎

## The Average of Loci Distances Over Nucleotide Sequences

As proposed earlier, the Genpofad distance can be extended from measuring distance between nucleotide sites to across an entire locus by taking an average.

$$\text{GENPOFAD}_{XY} = \frac{1}{n} \sum_{i=1}^{n} \text{GENPOFAD}_{XY}^{i}$$

**Theorem 9**. *Averages of GENPOFAD distances are also a metric.*

*Proof.* This average has symmetry by the fact that individual GENPOFAD distances are symmetric, and similarly, has the condition that the distance between a point is 0 if and only if the points are equal, by the fact that the averages of 0 are 0, and that GENPOFAD distances form a metric. Let there be 3 individuals, $A$, $B$, and $C$.

$$\overline{\text{GENPOFAD}(A, B)} \stackrel{\text{def}}{=} \frac{\text{GENPOFAD}_{A_1 B_1}^{i} + \ldots + \text{GENPOFAD}_{A_n B_n}^{i}}{n}$$

Since this is a scalar multiple of a sum of metrics, we have:

$$\begin{aligned}
\overline{\text{GENPOFAD}(A,B)} &\geq \frac{\text{GENPOFAD}^i_{A_1C_1} + \text{GENPOFAD}^i_{C_1B_1} + \ldots + \text{GENPOFAD}^i_{A_nC_n} + \text{GENPOFAD}^i_{C_nB_n}}{n} \\
&= \frac{\text{GENPOFAD}^i_{A_1C_1} + \ldots + \text{GENPOFAD}^i_{A_nC_n}}{n} + \frac{\text{GENPOFAD}^i_{C_1B_1} + \ldots + \text{GENPOFAD}^i_{C_nB_n}}{n} \\
&= \overline{\text{GENPOFAD}(A,C)} + \overline{\text{GENPOFAD}(C,B)}
\end{aligned}$$

The triangle inequality holds for the average of GENPOFAD distances. ∎

Genpofad$_{XY}$ will always take rational values between 0 and 1. As proven, this is also a metric. Treating these as random variables, since this is an average of independent and identically distributed variables, the expected value remains unchanged. The variance is altered; as we're dividing by $n$, the variance changes by a factor of $\frac{1}{n^2}$.

# Creating a Graph Representing a Population's Genetic Relationships at a single Locus

In this article, we used the stdpopsim tool by Adrion et al. , and in particular the Out of Africa model by Gutenkunst et al , modeling human genomic data as humans exanded out of Africa and into Europe and Asia. Our methodology is as follows: Given a sample of $n$ individuals, pulled from the same population, whose pairwise genetic distance at an individual locus has been computed using GENPOFAD. By choosing a "reasonable" cutoff value, again, see the discussion in subsection 3.3.1.

## Choice of Cutoff Value

We face a choice in what value to pick as a cutoff value for which edges between individuals are included. We started by looking over all the simulated individuals and seeing the distribution of distances between individuals within a population. For each population of 50 individuals, this will consist of $(50^2 - 50)/2 = 1225$ pairwise distances, across 45 populations, for a total of 55125 pairwise distances. The distribution of pairwise distances is shown in Figure 1. We observed our data exhibits a mean of 0.1365818 and a standard deviation of 0.005915615.

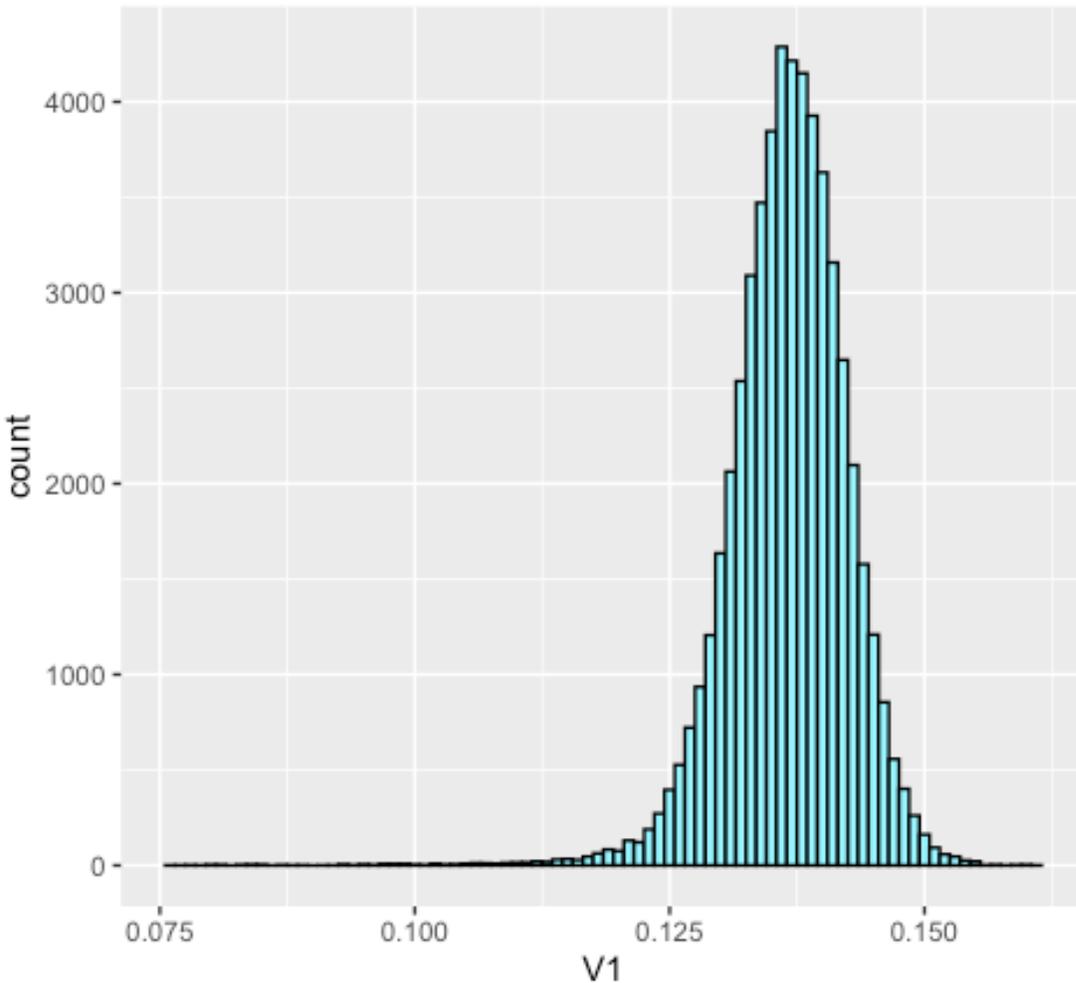

*Figure 1: Histogram showing the distribution of 55125 pairwise distances between simulated indivduals at a locust site of length*

We then considered what would be the smallest cutoff value to give us connected graphs. As discussed below, in order for a graph to be of small world type, it must be connected (that is, it must consist of one connected component). We began by an individual in the data and finding their minimum distance to a different individual. By taking the maximum value for this across each graph as a cutoff value, we would ensure that each vertex has degree at least 1. However, we note that this is not sufficient to ensure a fully connected graph, as a graph could consist of multiple connected components with high connectivity within the individual components, but the components are disconnected from each other. Next, we generated the mimimal spanning tree for each graph. A minimal spanning tree (MST) a subset of edges in a connected, edge-weighted graph that connects all vertices without any cycles and has the lowest possible total edge weight. It represents the most cost-effective way to link all the vertices in a graph. We generated each MST using Kruskal's algorithm . By generating the MST and then taking the maximum edge weight that appears in each MST, we are able to determine the minimum cutoff value that would generate a connected graph. The distributions of these values is shown in the histograms in Figure 2.

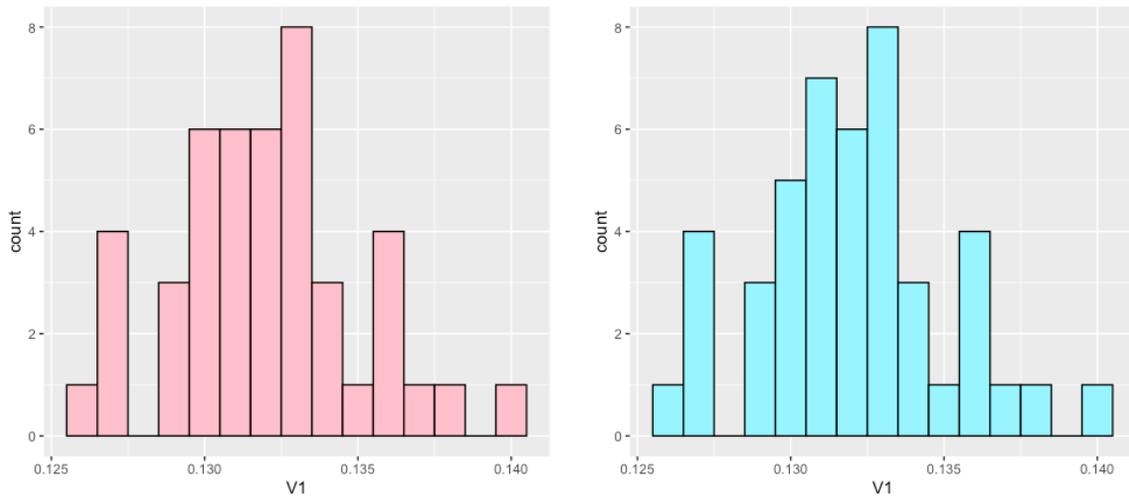

*Figure 2: Distribution of the value that would ensure each vertex of a population graph has degree at least 1 (left) and distribution of the minimum value that would guarantee each graph is connected (right), across a set of 45 population graphs.*

We note that the histograms in Figure 2 are very similar. In fact, the two data sets are identical except for two entries, where the values differ by a magnitude of $10^{-4}$. This suggests that when the graphs are disconnected, the last component to be connected is an isolated vertex. We agreed that an isolated vertex or individual may be a feature we did not want to exlude by forcing every graph to be connected. We noticed in our simulated data 7 of the graphs would remain disconnected if we used an umbrella cutoff value of 0.135, with a small gap in the data appearing at this point (a more thorough statistical analysis of this gap, and whether it is truly statistically significant, could be undertaken in future work). In fact, the data appears right skewed. We then settled on a global cutoff value of 0.135 in order to ensure most graphs are connected, while a few graphs do exhibit isolated vertices. We can then note the presence of these outlier individuals while continuing the analysis on the large connected component. With this value in mind, we note that our population graphs have an average edge density of 0.3589, to be used in Section 4.1.

### Adjacency Matrices

An adjacency matrix $A$ of an undirected graph G on $m$ nodes and $n$ edges is an $m \times m$ symmetric matrix where the $ij$'th element $a_{ij}$ either has entry 0 or 1. If the vertices $v_i$ and $v_j$ are adjacent (that is, there is an edge between them), $a_{ij}$ has entry 1. Otherwise, $a_{ij} = 0$. In a simple graph that does not allow for loops (an edge that connects a vertex to itself), these matrices have 0s along the diagonal.
Adjacency matrices are also useful for computing paths between two vertices. Given to vertices $v_i$ and $v_j$, the $ij$th entry of $A^n$ corresponds to the number of paths of length $n$ between $v_i$ and $v_j$.

## Cycles and Spanning Trees

Harary and Manvel found that there are multiple equations to find the number of "small" cycles in a graph. The presence of minimal cycles of length higher than 3 somewhat contradicts our idea of transitivity between individuals. Suppose that there are 3 individuals: individual A is (by our standards) genetically similar to both individuals B and C, but individuals B and C aren't genetically similar. This is expected to happen quite rarely compared to random graphs. From this, they have determined the number of 3, 4, and 5 cycles (shown respectively) for a graph $G$ with adjacency matrix $A$:

$$c_3(G) = \frac{1}{6}\text{tr}(A^3)$$

$$c_4(G) = \frac{1}{8}\left(\text{tr}(A^4) - 2|E(G)| - 2\sum_{i \neq j} a_{ij}^{(2)}\right)$$

$$c_5(G) = \frac{1}{10}\left(\text{tr}(A^5) - 5\text{tr}(A^3) - 5\sum_{i=1}^{|V(G)|}\left(\sum_{j=1}^{|V(G)|} a_{ij} - 2\right)a_{ii}^{(3)}\right)$$

# Statisical Testing on Population Graphs

We are interested in studying what behaviors these graphs exhibit in general. For example, it is reasonable to expect that exhibit a high number of three cycles and high clustering. This is due to an expected transitivity property: if individual $A$ is closely related to individual $B$, and individual $B$ is closely related to individual $C$, it stands to reason that individuals $A$ and $C$ would be closely related as well (though this property will weaken as the chain of individuals grows). In order to make this claim, we first compare population graphs with Erdős–Rényi graphs.

## The Erdős–Rényi Model

Erdős–Rényi graphs refers to two similar but distinct models for generating a random graph. The Erdős–Rényi model we will consider is given as follows: consider a graph $G$ on $n$ vertices. Over all $\binom{n}{2}$ possible edges in this graph, each edge will be included with probability $p$. The existence (or lack) of edges is pairwise independent. Such graphs on $n$ nodes and probability $p$ are denoted with the notation: $G(n,p)$. This model is also sometimes referred to as the Erdős–Rényi -Gilbert model.
It is well known that the number of 3-cycles in an Erdős–Rényi graph is not binomially distributed. This is because the probabilities of two 3-cycles appearing in a graph are not independent. Consider vertices $v_1, v_2, v_3, v_4$, and suppose an Erdős–Rényi graph $G(n,p)$ exhibits a 3-cycle on vertices $v_1, v_2, v_3$. Now consider

$$P(v_2, v_3, v_4 \text{ form a 3-cycle}) = p^3.$$

However,

$$P(v_2, v_3, v_4 \text{ form a 3-cycle} | v_1, v_2, v_3 \text{ form a 3-cycle}) = p^2$$

since, if $v_1, v_2, v_3$ form a 3-cycle, we know there is already an existing edge from $v_2$ to $v_3$. Therefore, since the events of individual 3-cycles are not, in general, independent, the distribution on the number of 3-cycles exhibited by an Erdős–Rényi graph is not binomially distributed.

To understand the distribution of 3-cycles on Erdős–Rényi graphs, we simulated 1,000,000 Erdős–Rényi graphs on 50 vertices with $p = 0.14$. The histogram on the number of three cycles in each graph is shown in Figure 3. It is known that despite the number of 3-cycles not being binomially distributed, the expected value of the number of three-cycles is given by

$$E[c_3] = \binom{n}{3} p^3.$$

On an Erdős–Rényi graph on 50 vertices with $p = 0.3589$, this number is given by

$$\binom{50}{3}(0.3589)^3 = 906.100659192$$

.

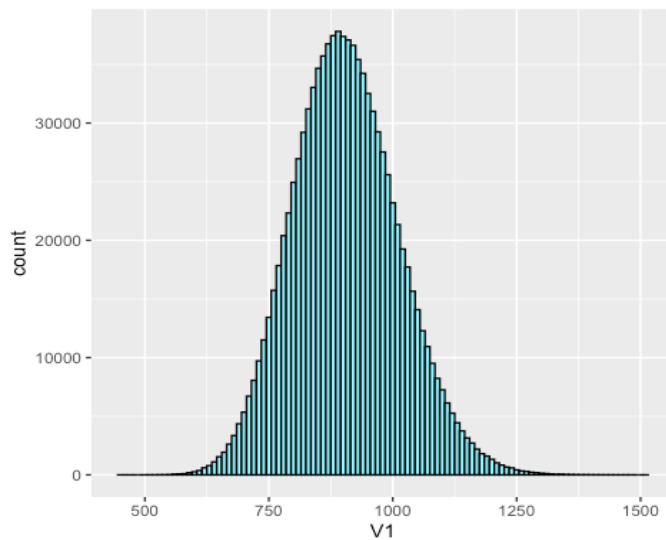

Figure 3: The simulated distribution of the number of 3-cycles in an Erdős–Rényi graph on from a simulation of one million graphs, generated on 50 vertices with $p = 0.3589$

Indeed, we note that in our set of 1,000,000 randomly simulated graphs, we see the sample average of the number of three cycles is $\bar{x}_{\text{ER}} \approx 906.1633$.

In addition, it is well-known that if $\varepsilon = \varepsilon(n)$ is any function such that $\varepsilon \to 0$ as $n \to \infty$, and if $p \geq \frac{\log n + \varepsilon}{n}$ then the Erdős–Rényi random graph $G(n,p)$ is asymptotically almost surely connected. In fact, in our simulation of 1,000,000 Erdős–Rényi graphs on 50 vertices with $p = 0.3589$, every graph generated was fully connected.

## Small-World Graphs

While Erdős–Rényi graphs form an intuitive model for random graphs, oftentimes graphs that represent real world data exhibit structure that one does not expect to find in an Erdős–Rényi graph. Many networks that represent real data from economic, biological, technological, or medical data instead falls somewhere between being Erdős–Rényi and regular. In , Watts and Strogatz describe a method to generate what are termed "small world" graphs: graphs that exhibit a high degree of clustering, and low shortest path lengths between nodes. By clustering, we mean the tendency of nodes to fall into highly connected subsets; the probability that two nodes in such a subset are connected by an edge is higher than two randomly selected nodes of the graph.

In , Humphries and Gurney define metrics to measure the "small-world-ness" of a network.

**Definition 10**. *A triplet is three vertices that are connected by either two (open) or three (closed) edges. A triangle graph then contains three closed triplets, one centered at each vertex.*

**Definition 11**. *The global clustering coefficient of a graph $G$ is defined as:*

$$C_G^\Delta = \frac{\text{number of closed triplets}}{\text{number of all triplets}} = \frac{3 \times \text{number of 3-cycles}}{\text{number of all triplets}}$$

They then define two new measures. The first,

$$\gamma_g^\Delta = \frac{C_g^\Delta}{C_{\text{Erdős–Rényi}}^\Delta},$$

is a ratio of the clustering coefficients between the graph of interest and an Erdős–Rényi graph on the same number of nodes *and* edges.

The second measure,

$$\lambda_g = \frac{L_g}{L_{\text{Erdős-Rényi}}},$$

corresponds to the ratio of $L_g$ to $L_{\text{Erdős–Rényi}}$. $L_i$ is found by taking the mean of the shortest path length of all combinations of any two nodes in graph $g$. This is known as the characteristic path length. Note then that it is a necessary condition that a graph be connected in order to be small world. If not, $\lambda_g = \infty$.

A graph is considered to be **small-world** if $\rho > 1$, where

$$\rho = \frac{\gamma_g^\Delta}{\lambda_g}.$$

It is well known (see ) that the expected global clustering coefficient of a Erdős–Rényi graph $G(n,p)$ is given by $E[C_{\text{Erdős–Rényi}}^\Delta] = p$ and the average shortest path length is given approximately by $L_{\text{Erdős–Rényi}} \approx \frac{\log(n)}{\log(p(n-1))}$. For $G(50, 0.3589)$ this value is $L_{\text{Erdős–Rényi}} \approx 1.3549$.

## Population Graphs

The number of three cycles in population graphs is expected to be significantly larger than the number of three cycles on an Erdős–Rényi graph on the same number of nodes and a similar number of edges. Unfortunately, as population graphs are more difficult to construct, the sample size is much smaller. For 43 simulated population graphs, the sample average of the number of three cycles is $\bar{x}_{\text{PG}} \approx 1677.07$. When we run a t-test against a null hypothesis that $\mu = 906.1$, we obtained a p-value of 0.001431, suggesting that the number of 3-cycles demonstrated by population graphs is significantly more than those demonstrated by Erdős–Rényi graphs. The 95% confidence interval is given by (1221.284, 2132.856) for the mean number of three-cycles exhibited by a population graph. Again, this is in line with our expectations as we would expect there to be some interdependence: if individual $A$ is closely related to individual $B$ and individual $B$ is closely related to individual $C$, there is a higher likelihood that individuals $A$ and $C$ are themselves closely related.

In fact, as mentioned above, there is reason to believe that the population graphs would demonstrate small world qualities. We noted that of our 43 generated population graphs, 7 were disconnected (as in, did not consist of one connected component). As discussed in section 3.3.1, this was usually due to one individual who had a degree of zero. This means these graphs are automatically not small world, as it is a necessary condition that small world graphs be connected. With this in mind, we then calculated the small world statistic $\rho$ for our 43 simulation populations graphs on their largest connected component. The histogram of these values can be found in Figure 4.

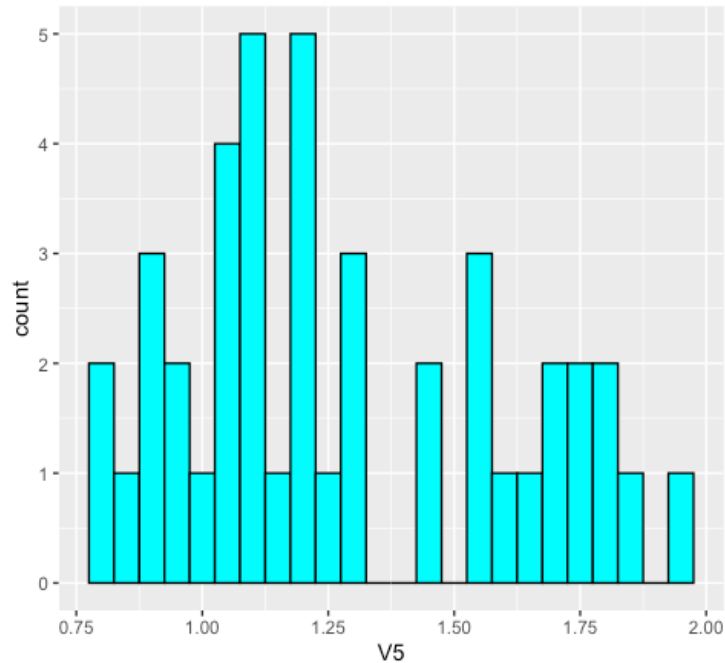

*The small world statistics ρ for 43 generated population graphs on their most connected component.*

The largest connected component of each of our 43 population graphs demonstrated an average value of $\hat{\rho} = 1.2819$, with an 95% confidence interval of $(1.1824, 1.3815)$. However, we notice that of these 43 graphs, 9 had values of $\rho$ less than 1, suggesting that they are not small-world. In fact, when considering the population graphs as a whole, there were a total of 14 graphs that were not small world, either due to being disconnected or due to having a value of $\rho$ less than 1. This seems like a rather large percentage, and a proportion test gives the 95% percent confidence interval of the proportion of population graphs that are not small world at $(0.1954, 0.4866)$. Thus, it appears that population graphs for our simulated populations, while on average do demonstrate small world properties, appear to frequently be distinct from small world graphs.

## Conclusion

The GENPOFAD distance relation is a useful one to determine genetic similarity between individuals with cutoff values. Using these cutoff values, "population graphs" and their respective structures can be studied. Particularly, the number of three cycles are more apparent, both intuitively and evidently. This, in part, supports the idea of transitivity: If individuals A and B are similar, and individuals B and C are similar, it can be concluded that individuals A and C are likely similar. This was proven with the comparison to random graphs on the same edge density.

With the introduction of small-world graphs, it has been found that on average, population graphs exhibit small-world properties. Investigating the properties of particular populations with population graphs which were not small-world could yield particularly interesting results on the population itself. To expand, the small-worldness of population graphs could be an indicator of unique properties to the population. Future work could expand on how exactly these graphs seem to differ from small world graphs and why. Future work could also be done to more closely simulate data from different types of populations to explore differences in the structures that comprise the population graphs.

## Acknowledgements

The authors would like to acknowledge Ali Al Balushi, Iyanna Penigo, Dr. Andre Kundgen, Dr. Mike Picollelli, and Dr. Arun Sethuramun for their counsel and advice during this work. This work was supported by the National Institute Health (Grant number 1R15GM143700-01).